\documentclass[reprint,aps,amsmath,amssymb,superscriptaddress]{revtex4-1}
\usepackage{graphicx}
\usepackage{dcolumn}
\usepackage{bm}
\usepackage{color,hyperref}

\newcommand{\fuzhou}{Department of Physics, Fuzhou University, Fuzhou 350108, Fujian, China }
\newcommand{\fujian}{Fujian Science and Technology Innovation Laboratory for Optoelectronic Information of China, Fuzhou 350108, Fujian, China}
\newcommand{\Italy}{Consiglio Nazionale delle Ricerche, Istituto dei Sistemi Complessi, via Madonna del Piano 10, I-50019 Sesto Fiorentino, Italy}
\newcommand{\italy}{Istituto Nazionale di Fisica Nucleare, Sezione di Firenze, via G. Sansone 1, I-50019 Sesto Fiorentino, Italy}

\begin{document}




\title{Heat conduction in low-dimensional electron gases without and with a magnetic field }


\author{Rongxiang Luo}
\email[]{phyluorx@fzu.edu.cn}
\affiliation{\fuzhou}
\affiliation{\fujian}
 \author{Qiyuan Zhang}
 \affiliation{\fuzhou}
 \author{Guanming Lin}
 \affiliation{\fuzhou}
\author{Stefano Lepri}
\email[]{stefano.lepri@isc.cnr.it}
\affiliation{\Italy}
\affiliation{\italy}


\date{\today}

\begin{abstract}

We investigate the behavior of heat conduction in two-dimensional (2D)
electron gases without and with a magnetic field.  We perform simulations with the Multi-Particle-Collision approach, suitably adapted to account for the Lorenz force acting on
the particles. For zero magnetic field, we find that the heat conductivity $\kappa$ diverges with the system size $L$ following the logarithmic relation $\kappa\thicksim \ln L$ (as predicted for two-dimensional (2D) systems) for small $L$ values; however, in the thermodynamic limit the heat conductivity tends to follow the relation $\kappa\thicksim L^{1/3}$, as predicted for one-dimensional (1D) fluids. This suggests the presence of a dimensional-crossover effect in heat conduction in electronic systems that adhere to standard momentum conservation.
Under the magnetic field, time-reversal symmetry is broken and the standard momentum conservation in the system is no longer satisfied but the \emph{pseudomomentum} of the system is still conserved.
In contrast with the zero-field case, both equilibrium and non-equilibrium
simulations indicate a finite heat conductivity independent on the system size $L$ as $L$ increases. This indicates that  pseudomomentum conservation
can exhibit normal diffusive heat transport,  which differs from the abnormal behavior observed in low-dimensional coupled charged harmonic oscillators with pseudomomentum conservation in a magnetic field. These findings support the validity of the hydrodynamic theory in electron gases and clarify that pseudomomentum conservation is not enough to ensure the anomalous behavior of heat conduction.
\end{abstract}


\maketitle

\section{Introduction\label{sec1}}

Insights on the microscopic mechanisms  of energy transport are crucial to achieve understanding of macroscopic irreversible heat transfer \cite{Lepri2016,Dhar2008,Lepri2003}. Also it serves as a theoretical foundation for thermal energy control and management~\cite{Li2012,Zhang2020,Gu2018,Maldovan2013}.
This is even more relevant at the nano- and microscale where novel effects
caused by reduced dimensionality, disorder and nano-structuring
affects natural and artificial
materials \cite{benenti2023non} .
Macroscopic behavior of thermal transport in the linear response regime is governed by well-known Fourier's law,
\begin{equation}\label{Eq1}
  j=-\kappa \nabla T,
\end{equation}
where $j$ and $\nabla T$ are, respectively, the heat current and the spatial temperature gradient, and $\kappa$ is the thermal conductivity, a finite constant independent of the system size. The transport following this law is usually known as normal heat conduction and it has been shown that as long as some nonlinearity is present in the interaction, heat conduction is normal for momentum-nonconserving systems~\cite{Lepri2016,Dhar2008,Lepri2003} (except for the coupled rotor systems~\cite{Giardina2000,Gendelman2000}), and for three-dimensional systems~\cite{Wang2010,Saito2010,Luo2021}.\par

Interestingly, heat conduction generically exhibits the different abnormal behaviors for low-dimensional systems with a conserved momentum. Specifically, for 2D momentum-conserving systems, it is expected from different theoretical apporaches~\cite{Lepri2002,Narayan2002} and also from an exactly solved stochastic model~\cite{Basile2006}, that $\kappa$ diverges logarithmically with the system size $L$,   $\kappa\sim \ln L$.  So far, such prediction has been numerically verified in 2D lattices~\cite{Lippi2000,Delfini2005,Yang2006,Xiong2010}, and only
recently in 2D fluids~\cite{Cintio2017,Luo2020}. On the other hand, for 1D momentum-conserving systems, $\kappa$ usually diverges as a power-law: $\kappa\sim L^{\alpha}$ \cite{Lepri1997}. The exponent $\alpha=1/3$ is related to the fact that fluctuations of the conserved field belong to the universality class of the famous Kardar-Parisi-Zhang (KPZ) equation~\cite{Beijeren2012,Spohn2014}. As for the 2D case, this prediction has been well numerically verified in several works both for lattices~\cite{Das2014,Wang2016} and fluids~\cite{Dhar2001,Casati2003,Chen2014}. Moreover, it has been found that in 2D lattice~\cite{Zolotarevskiy2015,Wang2020} and fluid~\cite{Cintio2017,Luo2020} systems with a conserved momentum, heat conduction under certain conditions can present the 2D to 1D dimensional-crossover effect. To summarize, under the same momentum conservation condition, heat transport can exhibit the same behaviors in lattices and fluids.\par

Another interesting issue that has been addressed only recently concern the effects of a magnetic field.
Some  results were obtained for heat transport in harmonic chains made of charged oscillators that interact with an external magnetic field~\cite{Tamaki2017,Saito2018,Tamaki2018,Bhat2022}. In particular, heat transport via the one-dimensional charged particle systems with transverse motions is studied in~\cite{Tamaki2017}, where researchers studied two cases: case (I) with uniform charge and case (II) with alternate charge. Due to the presence of the magnetic field, the standard momentum conservation in such a system is no longer satisfied but is replaced by the pseudomomentum conservation. An interesting result found in this work is that for both cases with nonzero magnetic field, heat conduction behaviors are abnormal, which is same as that of zero magnetic field. Remarkably, the divergence exponent $\alpha$ may be different from the
case without field, suggesting a novel dynamical universality class. This finding is at variance with that obtained in the momentum-nonconserving systems~\cite{Lepri2016,Dhar2008}, and thus, we raise the relevant question: Does the pseudomomentum conservation of a system leads to anomalous heat conduction? More interestingly, in case (I) the anomalous heat conduction with a new exponent is theoretically and numerically demonstrated, exhibiting a diverging thermal conductivity with vanishing sound speed. To verify whether this new anomalous behavior of heat conduction apply equally well to both lattices and fluids, the research evidence for the counterpart fluids is in urgent need at present. This then raises another question: Can the anomalous thermal conductivity with a new exponent also be observed in low-dimensional fluids under the same pseudomomentum conservation?\par

In this work, we give a clear answer to the above two questions. To this aim, we consider a low-dimensional system of  charged particles in a rectangular box, interacting through the multi-particle collision (MPC) dynamics~\cite{Malevanets1999}. This choice is particularly  convenient
since it  correctly capture the hydrodynamic equations with great advantages in the numerical simulations~\cite{Padding2006}. By using the MPC dynamics, researchers achieved a considerable understanding of various aspects of transport~\cite{Gompper2009,Benenti2014,Luo2018}. Importantly, the MPC dynamics conserves the total momentum and energy of the system, and thus it can help us test the theoretical conjecture for momentum-conserving systems~\cite{Cintio2017,Luo2020}.\par

Specifically,  we numerically study heat transport properties of electron gases without and with a magnetic field both by equilibrium (Green-Kubo)  and nonequilibrium (thermal-wall) methods. We first demonstrate that for zero magnetic field, the system with momentum conservation   exhibits the crossover from 2D to 1D behavior of the thermal conductivity.  Then we show that low-dimensional electron gases with the pseudomomentum conservation in a magnetic field can exhibit normal heat conduction behavior. This is different from the abnormal behavior observed in
the other class of models,  low-dimensional coupled charged harmonic oscillators mentioned above. Our results thus clarify that the pseudomomentum conservation is not related to the normal and anomalous behaviors of heat conduction and provide an example of the difference in heat conduction between fluids and lattices in the presence of the magnetic field. \par

\section{The low-dimensional electron gas model\label{sec2}}
\begin{figure}
  \centering
  \includegraphics[width=8cm]{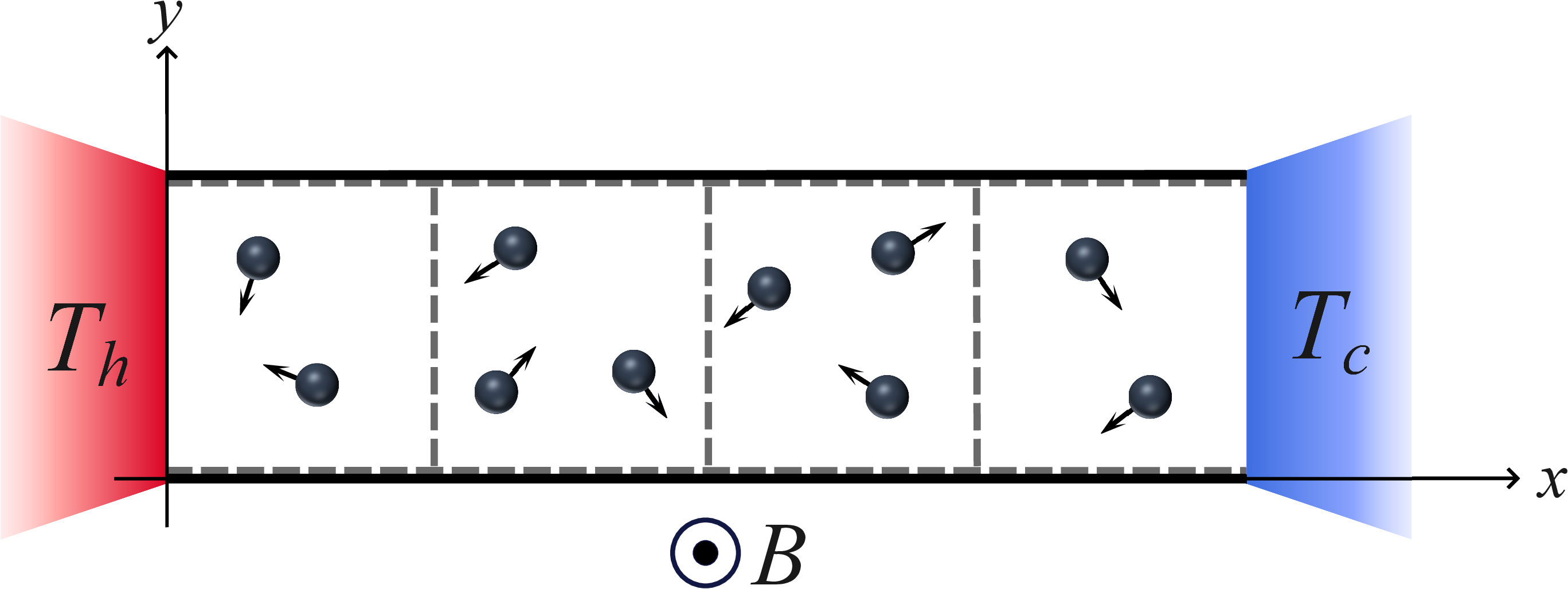}\\
  \caption{Schematic plot of the low-dimensional electron gas system in a rectangular box of width $W$ and length $L$ evolving with the MPC dynamics. A magnetic field, transverse to the plane of motion, is applied to the system which is coupled to two heat baths at temperatures $T_{h}$ and $T_{c}$. The cells of dashed-line boundaries represent the partition of space considered for modeling collisions. The $x$ coordinate goes along the channel and $y$ is perpendicular to it. See text for the more details. }\label{fig1}
\end{figure}

We consider an ensemble of  $N$ interacting point particles, of mass $m_i$ and charge $e_i$, where $i=1, \ldots, N$. All particles are in a rectangular box of length $L$ (along the $x$ coordinate) and width $W$ (along the $y$ coordinate) (see Fig.~\ref{fig1} for a schematic plot). A constant magnetic field  perpendicular to the plane of motion, $\textbf{\emph{B}}=B\textbf{\emph{k}}$, is applied to the system. In the $y$-direction the particles are subject to periodic boundary conditions. In the non-equilibrium setup, the system is placed in contact with two heat baths at $x=0$ and $x=L$, through openings of the same size as the width $W$ of the box. The heat and cold baths are modeled as ideal gases and characterized by two different temperatures $T_h$ and $T_c$, respectively.
We implement this through the thermal wall approach:
whenever a particle hits the boundaries of the system, it is reflected back with a new velocity (denoted by $v_x$ and $v_y$ of its $x$ and $y$ components, respectively) randomly drawn from a Maxwellian distribution with probability densities~\cite{Lebowitz1978}
\begin{equation}\label{Eq2}
\begin{aligned}
  f\left(v_{x}\right)= & \frac{ m|v_{x}|}{k_{B}T_\iota}\textrm{exp}\left(-\frac{mv^{2}_{x}}{2k_{B}T_\iota}\right), \\
  f(v_{y})= &\sqrt{\frac{m}{2\pi k_{B}T_\iota}}\textrm{exp}\left(-\frac{mv^{2}_{y}}{2k_{B}T_\iota}\right),
\end{aligned}
\end{equation}
where $T_\iota$ ($\iota=h, c$) is the temperature of the respective heat bath in dimensionless units, $k_{B}$ is the Boltzmann constant.\par

The dynamics in the bulk is described by the MPC dynamics.
Loosely speaking, the MPC simplifies the numerical simulation of interacting particles by coarse-graining time and space
and replacing deterministic interaction forces with random ones. The MPC dynamics assumes that the system evolves in discrete time steps, consisting of the propagation dominated only by the Lorentz force during a time $\tau$ followed by instantaneous collision events. During the propagation period, the evolution equations for this system are~\cite{Bonella2014}
\begin{equation}\label{Eq3}
\begin{aligned}
      \dot{x}_i= &\frac{p_{i}^x}{m_i}+\omega_i y_i, \\
      \dot{y}_i= & \frac{p_{i}^y}{m_i}-\omega_i x_i ,\\
  \dot{p}_{i}^x= &\omega_i\left(p_{i}^y-m_i \omega_i x_i\right) ,\\
  \dot{p}_{i}^y= -&\omega_i\left(p_{i}^x+m_i \omega_i y_i\right) ,
\end{aligned}
\end{equation}
where $\textbf{\emph{p}}_{i}=m_i\textbf{\emph{v}}_{i}$ and $\omega_i=\frac{e_iB}{m_i}$ are the momenta and cyclotron frequence of particle $i$. For each instantaneous collision event, the system's volume is partitioned into identical square cells of size $a \times a$ (see Fig.~\ref{fig1}), and then the velocities of all particles found in the same cell are rotated with respect to their center of mass velocity $\textbf{\emph{V}}_{\textmd{c.m.}}$ by two angles, $\alpha$ or $-\alpha$, randomly chosen with equal probability. The velocity of a particle in a cell is thus updated as
\begin{equation}\label{Eq4}
 \textbf{\emph{v}}_i\rightarrow\textbf{\emph{V}}_{\textmd{c.m.}}\\
+\hat{\mathcal{R}}^{\pm\alpha}\left(\textbf{\emph{v}}_i-\textbf{\emph{V}}_{\textmd{c.m.}}\right),
\end{equation}
where $\hat{\mathcal{R}}^{\pm\alpha}$ is the 2D rotation operator.  The time interval between successive collisions $\tau$ and the collision angle $\alpha$ tune the strength of the interactions and consequently affect the transport of the particles. Note that the angle $\alpha=\pi/2$ corresponds to the most efficient mixing of the particle momenta and such MPC dynamics preserve the total momentum and  energy of the electron gas system.\par

To compare with the results obtained in Ref.\cite{Tamaki2017}, we also consider two cases: case (I) with uniform charges $e_i=1$ and case (II) with opposite charges
on each half of particles,  say $e_i=(-1)^i$ . We find that from the dynamics described above, the sum of the following variable is conserved~\cite{Tamaki2017}:
\begin{equation}\label{Eq5}
\textbf{\emph{P}}_i= \textbf{\emph{p}}_i-e_i \textbf{\emph{r}}_i\times \textbf{\emph{B}},
\end{equation}
where vector $\textbf{\emph{r}}_i(x_i,y_i)$ specifies the position of the $i$th particle. The variable $\textbf{\emph{P}}_i$ is a pseudomomentum which is not equivalent to the canonical momentum $\textbf{\emph{p}}_i$. Hence, when a magnetic field is applied, breaking the time-reversal symmetry, the standard momentum conservation  is replaced by the pseudomomentum conservation. We should point out that to keep pseudomomentum conserved,  periodic boundary conditions are applied to the system only in  instantaneous collision events. Actually, the sum of $\textbf{\emph{P}}_i$ is conserved, which is generally the case for Hamiltonian systems in a magnetic field~\cite{Johnson1983}. \par

\begin{figure}
\includegraphics[width=9cm]{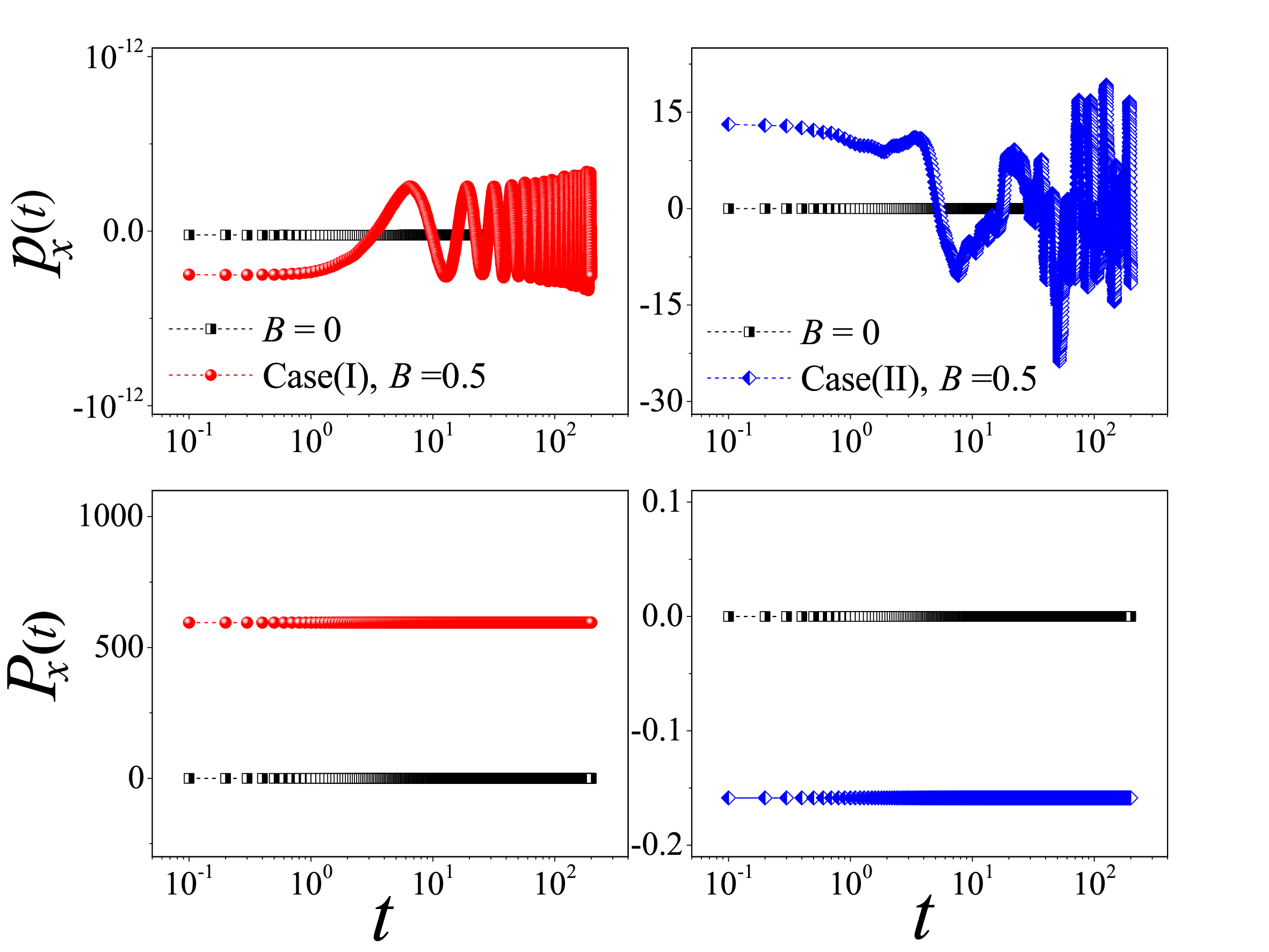}
\caption{ The total momentum $p_x(t)$ and total pseudomomentum $P_x(t)$ in the $x$ direction as a function of the evolution time $t$ for the zero magnetic field case $(B=0)$ and the nonzero magnetic field case $(B=0.5)$. Here, we set $L=32$.}
\label{fig2}
\end{figure}

In our simulations,  we initially assign to each particle a random position uniformly distributed and a random velocity drawn from the Maxwellian distribution at a nominal temperature $T$. We set $T_\iota$ to be slightly biased from $T$, i.e., $T_{h,c}=T\pm\Delta T/2$, to measure the temperature profile $T(x)$, where $x$ is the space variable, and the heat current $j$. Numerically, $T(x)$ and $j$ are obtained in a same way as in the 2D case~\cite{Luo2020}. The heat conductivity is finally obtained, by assuming the Fourier law, as $\kappa\approx jL/(T_h-T_c)$, where we have checked that in the linear response regime, the temperature jump between the heat bath and simulated system is not sensitive to the choice of $L$ and thus can be neglected (as shown in Fig.~\ref{fig6} of this paper). We set the main parameters as follows: $T=k_B=m_i=1$, $\Delta T=0.2$, $W=a=\tau=0.1$, $\theta=\pi/2$, and the averaged particle number density $\rho=N/(LW)=22$. In this work, long enough integration times $\left(>10^8\right)$ are utilized to ensure the relative errors of all numerical results are smaller than 0.5$\%$ .\par

\section{Equilibrium simulations results \label{sec3}}
Now let us turn to the simulations results. In the equilibrium simulations, we consider an isolated system with periodic boundary conditions also in the $x$ direction. The initial condition is randomly assigned with the constraints that the total momentum is zero and the total energy corresponds to $T=1$. The system is then evolved and after the equilibrium state is attained, we compute the transport properties of the system. \par

First of all, we examine the total momentum conservation and  total pseudomomentum conservation of the system when the magnetic field is absent and when it is present. It can be seen in Fig.~\ref{fig2} that for $B=0.5$, the total momentum in the $x$ direction $p_x(t)$ for case(I) and case(II), when compared to $B=0$, changes noticeably with the evolution time $t$. On the other hand, we can see that the total pseudomomentum $P_x(t)$ remains constant over time for both cases with $B=0.5$. These numerical evidences verify that when the time-reversal symmetry is broken by an applied magnetic field, the standard momentum conservation of the system is replaced by its pseudomomentum conservation.\par

\begin{figure}
\includegraphics[width=8.5cm]{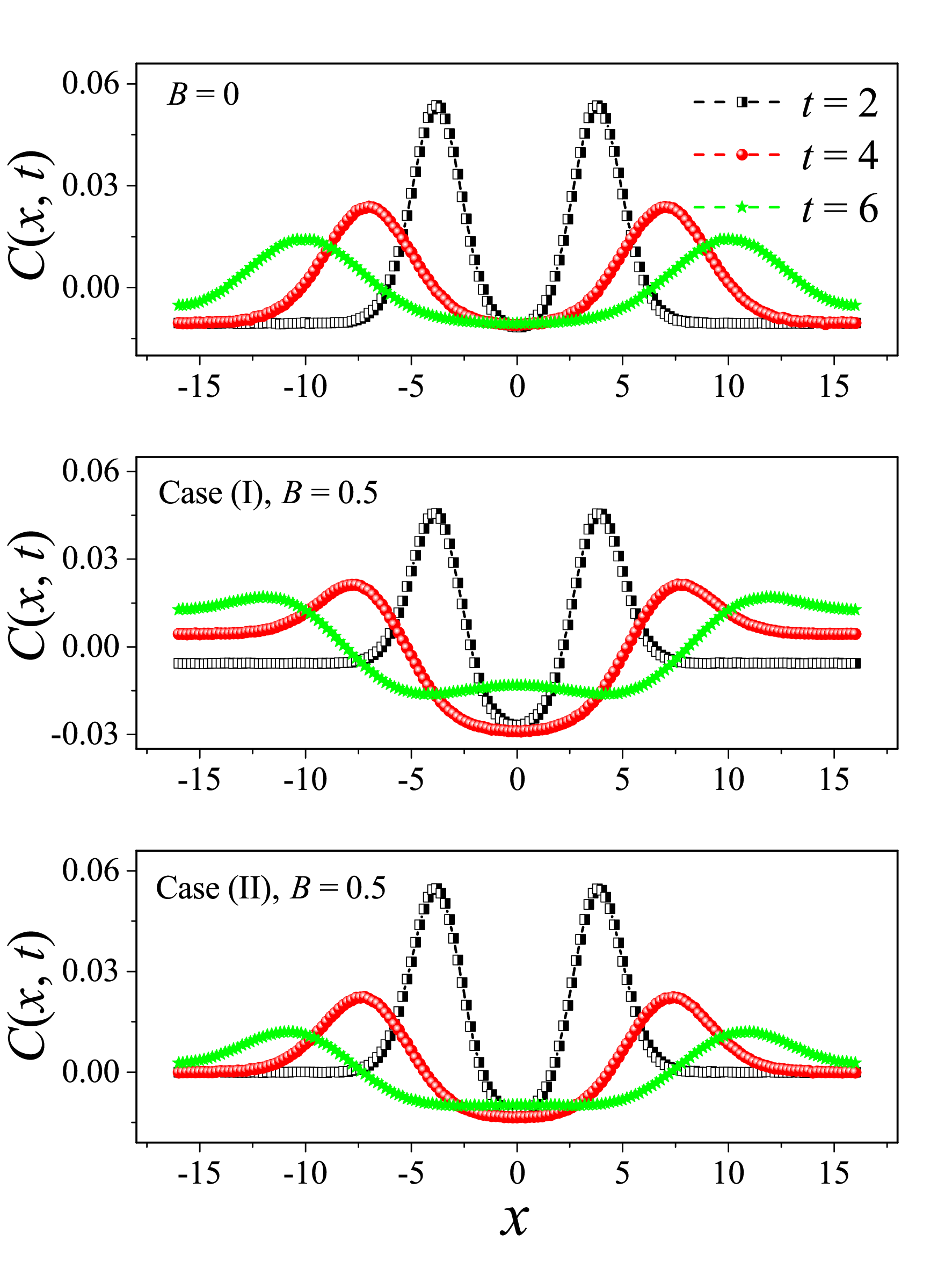}
\caption{ Numerical calculation of the spatiotemporal correlation function $C\left(x,t\right)$ for the electron gases without and with a magnetic field. Here the system length is set to be $L=32$. One can clearly see the two peaks (the hydrodynamic mode of sound) moving oppositely away from $x=0$ in $B=0$ and $B=0.5$ for case (I) and case (II).}
\label{fig3}
\end{figure}

As performed in~\cite{Tamaki2017}, we next resort to the spatiotemporal correlation function of local heat currents to check the sound velocity of the system. The spatiotemporal correlation function of local heat currents is defined as~\cite{Casati2003,Zhao2006,Levashov2011}
\begin{equation}\label{Eq5}
C\left(x,t\right)\equiv \left<J^{loc}(0,0)J^{loc}(x,t)\right>.
\end{equation}
Numerically, we compute $C\left(x,t\right)$ as performed in~\cite{Zhang2014}: the system is divided into $\frac{L}{b}$ bins in space of equal width $b=0.2$; the local heat current in the $k$\textmd{th} bin and at time $t$ is defined as $J^{loc}(x,t)\equiv\sum_{i}\frac{1}{2}m_i\textbf{\emph{v}}^{2}_iv_{x,i}$, where $x\equiv kb$ and the summation is taken over all particles that reside in the $k$\textmd{th} bin.
It is  found that $C\left(x,t\right)$ features a pair of pulses moving oppositely away from $x=0$ at the sound speed~\cite{Prosen2005,Lepri1998,Chen2013}, which are recognized to be the hydrodynamics mode of sound. In Fig.~\ref{fig3}, we present $C\left(x,t\right)$ for the system size $L=32$. The two peaks representing the sound mode can be clearly identified in Fig.~\ref{fig3}(a), (b), and (c). Their moving speed $v_{s}$ is measured to be $v_s\simeq1.55$ for $B=0$, $v_s\simeq2$ for case (I) with $B=0.5$, and $v_s\simeq1.8$ for case (II) with $B=0.5$. As mentioned in the Introduction, case (I) in~\cite{Tamaki2017} exhibits a diverging thermal conductivity with vanishing sound speed. Obviously, there is a difference for case (I) between lattices and fluids, indicating possibly different transport mechanisms. \par

Let us now  investigate the heat conduction behavior of the system by the celebrated Green-Kubo formula. This formula relates transport coefficients to the current time-correlation functions, and thus the heat conductivity can be expressed as~\cite{Dhar2008,Lepri2003,Kubo1991}
\begin{equation}\label{Eq6}
\kappa_{GK}=\frac{\rho}{k_BT^2}\lim_{\tau_{\mathrm{tr}}\rightarrow\infty}\lim_{N\rightarrow\infty}\frac{1}{N}\int_{0}^{\tau_{\mathrm{tr}}}\left<J(0)J(t)\right>dt.
\end{equation}
In this formula, $\left<J(0)J(t)\right>$ is the total heat current correlation function in the equilibrium state, where $J\equiv \Sigma_{i}^{N} \frac{1}{2}m_i\textbf{\emph{v}}^{2}_iv_{x,i}$ represents the total heat current along the $x$ coordinate. To calculate the heat conductivity, the integral is usually truncated up to $\tau_{\mathrm{tr}}=L/v_{s}$ ($v_{s}$ is the sound speed)~\cite{Dhar2008,Lepri2003}. This results the superdiffusive heat transport $\kappa_{GK}\sim L^{1-\lambda}$ as long as $\left<J(0)J(t)\right>$ decays as $\sim t^{-\lambda}$ with $\lambda <1$. \par

\begin{figure}
\includegraphics[width=9cm]{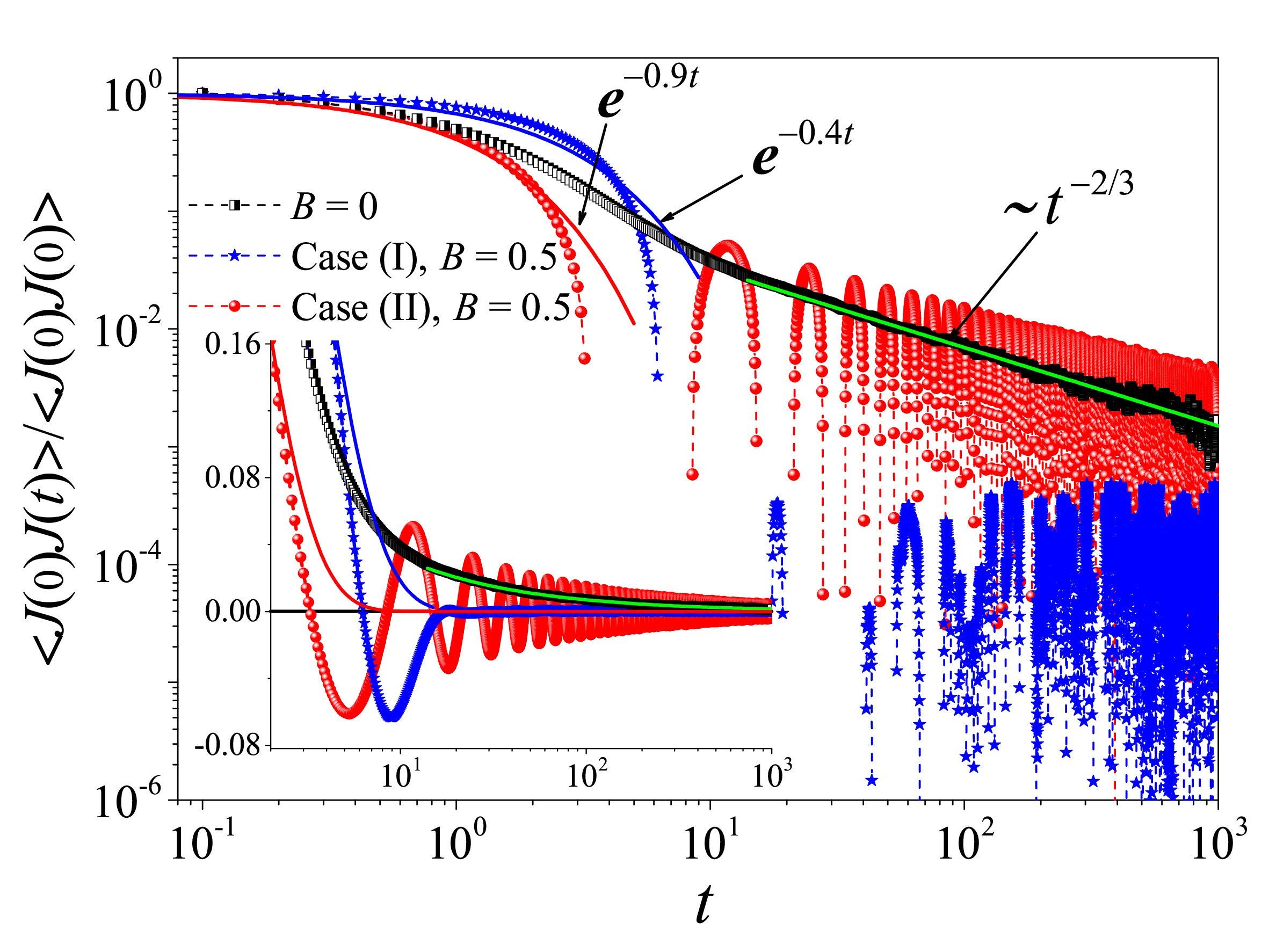}
\caption{ The total heat current auto-correlation function for the low-dimensional electron gases. The blue line and the red line are the fit of the data for both cases with $B=0.5$, and the green line at $B=0$ indicate the decay with $t$ as $\sim t^{-2/3}$. The inset: the log-linear scale is plotted to appreciate the oscillations around zero. Here we fix $L=2560$.}
\label{fig4}
\end{figure}

The results of $\left<J(0)J(t)\right>$ versus $t$ for $B=0$ and $B=0.5$ are presented in Fig.~\ref{fig4}. It can be seen that for $B=0$, after a rapid decay at short times, the correlation function eventually attains a power-law decay
$\left<J(0)J(t)\right>\sim t^{-\lambda}$ with $\lambda=2/3$, showing a long slowly decaying hydrodynamic tail. This result, combined with the Green-Kubo formula, leads to $\kappa_{GK}\sim L^{1/3}$, fully compatible with the theoretical prediction of the 1D case~\cite{Beijeren2012,Spohn2014}. However, it is clear that for $B=0.5$, the correlation function for both cases undergoes an exponential decay at short times, and eventually, when $t>4$, it begins to oscillate around zero (see the inset of
Fig.~\ref{fig4}).\par

A relevant question concerns the limit of a very weak field.
Upon increasing $B$ from zero, one may expect that the decay should occur on
faster and faster time scales.
Indeed, we have observed (data not reported)
that even for relatively small values, say $B=0.01-0.05$, the decay turns to
be exponential on the accessible simulation times. Only some minor differences
are seen for case (II) on intermediate time scales (e.g. $t<10^{3}$ for $B=0.01$)
where correlations decay slower. Altogether, we conclude that
no significant qualitative differences were observed in the results and that
an arbitrary small $B$ suffices to yield normal transport. This is reminiscent of the known fact that an arbitrarily weak force breaking momentum conservation restores Fourier's law for oscillator chains~\cite{Lepri2016,Dhar2008,Lepri2003}.\par

\section{Non-equilibrium simulations results \label{sec4}}
\begin{figure}
\includegraphics[width=9cm]{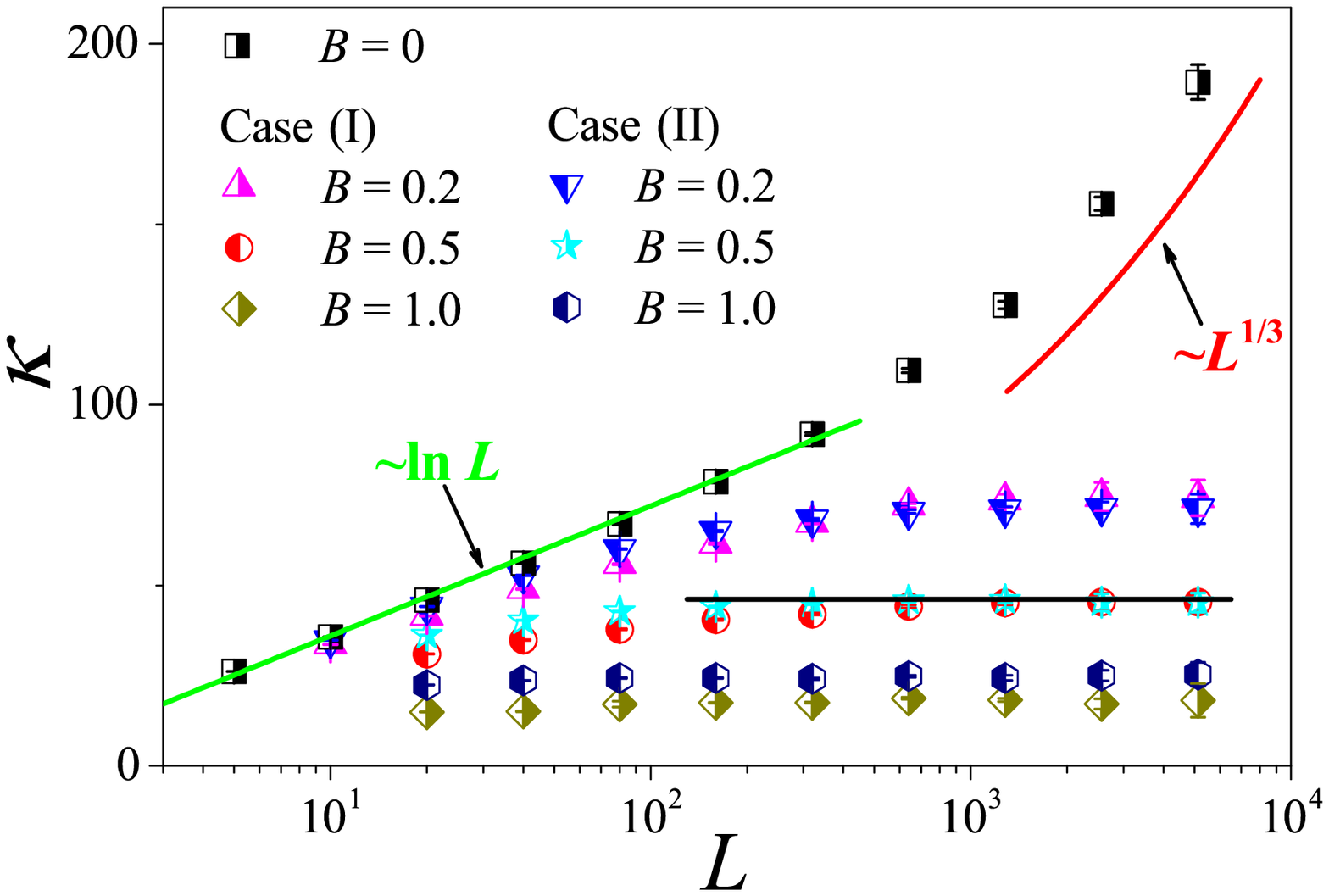}
\caption{The heat conductivity $\kappa$ as a function of the system length $L$ for the system without and with a magnetic field. The symbols are for the numerical results. For reference the green straight line is the best logarithmic fit, $\kappa\sim \ln L$ and the red curve line indicates the divergence with $L$ as $\sim L^{1/3}$. For $B=0.5$, the black horizontal line denotes the saturation value of $\kappa_{GK}$ obtained by Eq.~(\ref{Eq6}), where the integration is up a time $L/v_s$
with $v_s$ measured in Fig.~\ref{fig3}.}
\label{fig5}
\end{figure}

Next, let us further confirm the results obtained above by nonequilibrium thermal-wall method. Based on Fourier's law [Eq.~(\ref{Eq1})], we plot in Fig.~\ref{fig5} the relation of $\kappa$ vs $L$ for various $B$. It can be seen that for $B=0$, $\kappa$ perfectly diverges with $L$ as $\kappa\thicksim \ln {L}$ for a small range of $L$. This is clear evidence of superdiffusive heat conduction behavior following the logarithmic relation in a low-dimensional system with momentum conservation. This fast convergence to the 2D theoretical prediction of $\kappa\thicksim \ln {L}$ law suggests that the gas system is a ideal platform to check the validity of existing theories~\cite{Lepri2016,Basile2006}. Moreover, it can also be seen that for $B=0$, by further increasing $L$, $\kappa$ eventually approaches the scaling $\kappa\sim L^{1/3}$ predicted in 1D momentum-conserving fluids~\cite{Narayan2002,Beijeren2012,Spohn2014} and also recent experimental observations in 2D materials~\cite{Xu2014}. This result is no strange since by increasing $L$, the dimensionality of our system can be naturally changed from 2D to quasi-1D. It thus suggests that such dimensionality-crossover effects would be ubiquitous in low-dimensional systems and essentially root in the underlying hydrodynamics. \par

\begin{figure}
\includegraphics[width=8.5cm]{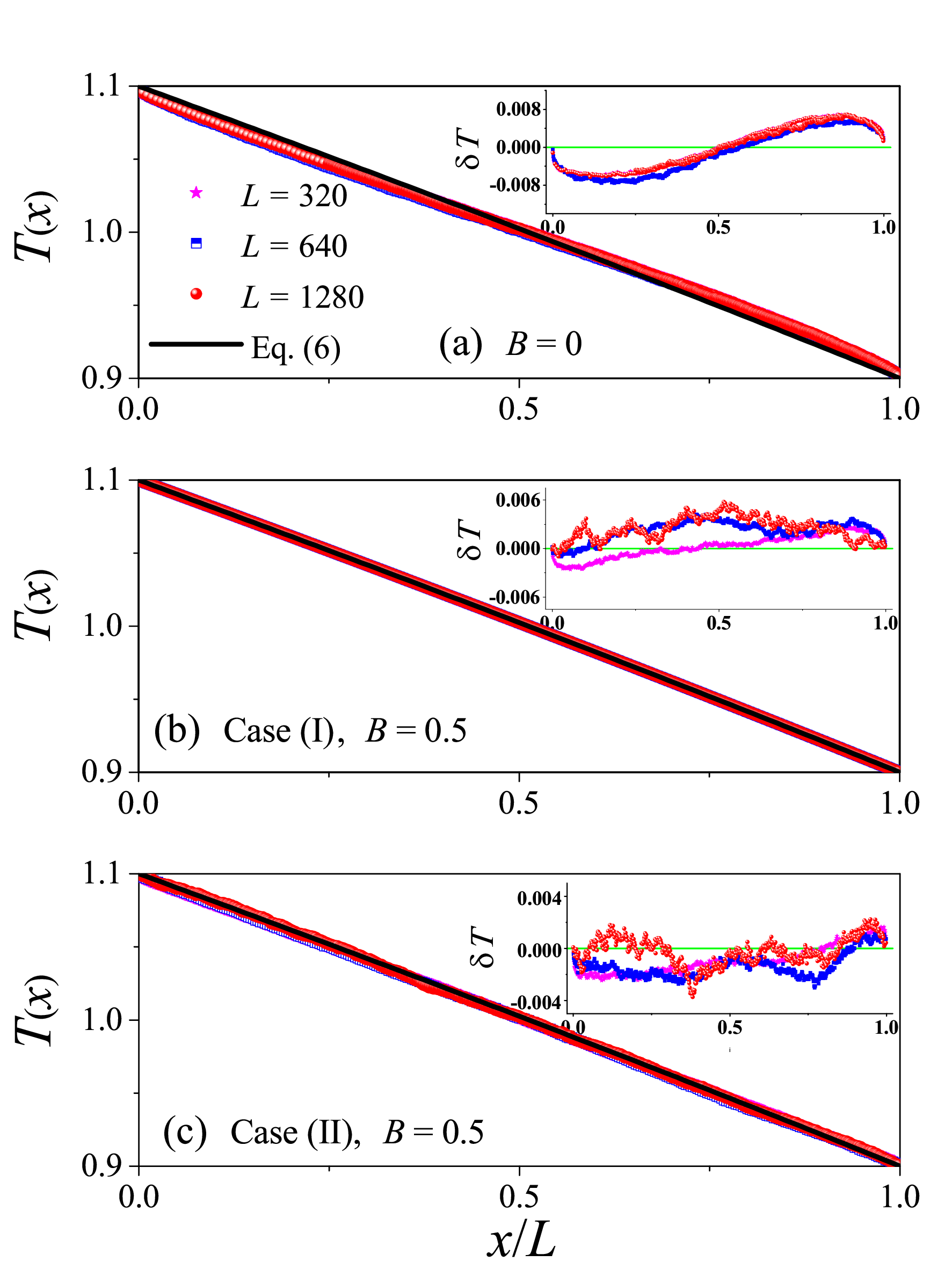}
\caption{Typical temperature profile $T(x)$ for the system without and with a magnetic field. Here, our numerical results are compared with the analytical expression [see Eq.~(\ref{Eq7})]. The inset: Plot of the differences of $\delta T$ between the data and the black line, and the green line at $\delta T=0$ are for reference.}
\label{fig6}
\end{figure}
However, we can see in Fig.~\ref{fig5} that for both cases with $B\neq 0$, as $L$ increases, $\kappa$ no longer diverges but rather approaches a finite value. Thus, the nonequilibrium simulations are fully consistent with the fast decay of equilibrium correlation.
%
Obviously, our results are at variance with the findings in~\cite{Tamaki2017}.
There, heat conduction in presence of pseudomomentum conservation in two cases are abnormal and even with a new diverging exponent in case (I). This observation, together with our results, thus provide evidence that the pseudomomentum conservation in the system is not related to the normal and anomalous behaviors of heat conduction. Finally, we also note that the value of finite heat conductivity decreases with an increase of $B$, as expected since the stronger the magnetic field $B$ is, the greater the heat resistance is.\par

The difference between normal and abnormal conduction can be further appreciated in the non-equilibrium steady-state temperature profiles $T(x)$. For systems with normal transport, $T(x)$  is determined by solving the stationary heat equation.
Assuming that the thermal conductivity is proportional to $\sqrt{T}$ as prescribed by standard kinetic theory, and imposing the boundary conditions, this yields~\cite{Dhar2001}
\begin{equation}\label{Eq7}
T(x)=\left[T_{L}^{3/2}\left(1-\frac{x}{L}\right)+T_{R}^{3/2}\frac{x}{L}\right]^{2/3}.
\end{equation}
In Fig.~\ref{fig3} this prediction (black line) is compared with our simulation results and to better appreciate the deviations from the prediction, we plot the differences $\delta T$ between the data and the black line in the inset. On the other hand, for systems with abnormal heat conduction $T(x)$ is expected to be qualitatively different, being solution of a fractional diffusion equation as demonstrated in several cases \cite{Lepri2011,Kundu2019}. A typical feature for abnormal behavior is that $T(x)$ is concave upwards in part of the system and concave downwards elsewhere, and this is true even for small temperature differences~\cite{Lippi2000,Mai2007,Lepri2009}. This is verified in the inset of Fig.~\ref{fig6}(a) by our numerical simulations for $B=0$ . Note that the data for three different $L$'s overlap with each other, implying that the deviations from Fourier's behavior are not finite-size effects. On the contrary, it is shown in the inset of Fig.~\ref{fig6}(b) and (c) that for both cases with $B=0.5$, $\delta T$ for various $L$ oscillates around zero, indicating that there is a good agreement between the results of our numerical simulations and Eq.~(\ref{Eq7}). Altogether, those numerical results again support our findings based on the length-dependence of the thermal conductivity, that heat conduction for the electron gases without a magnetic field is abnormal while with a magnetic field it is normal.

\section{Summary and discussions\label{sec5}}
We used  both equilibrium and nonequilibrium molecular dynamics methods to examine the heat conduction properties of the low-dimensional electron gases
with MPC dynamics,  without and with a magnetic field. Both approaches
give sound confirmation that  without  magnetic field there exist a crossover from 2D to 1D behavior of the thermal conductivity. This result is rooted in the condition for low-dimensional systems described by MPC dynamics to keep the momentum conservation of the system.  The fact that the decay exponent of the
current autocorrelation matches the theoretical prediction of fluctuating
hydrodynamics \cite{Spohn2014} supports its validity also for
quasi-1D systems.

We then provided a strong evidence that the Fourier law is
recovered in low-dimensional
MPC fluids when the momentum conservation is broken by the field, somehow
similar to what happens when momentum conservation is broken by an
external pinning potential~\cite{Lepri2016,Basile2006}. More importantly, our results suggests that under the same pseudomomentum conservation, the behavior of heat conduction in low-dimensional fluids differs from the abnormal behavior observed in chains of coupled charged harmonic oscillators~\cite{Tamaki2017}. In this respect, it appears
that the two classes of models behave differently as soon as an arbitrarily
weak field is switched on. This empirical observation calls for a theoretical
explanation.

This finding clarifies that pseudomomentum conservation is not enough to ensure the anomalous behavior of heat conduction and provides an example of the difference in heat conduction between fluids and lattices in the presence of the magnetic field condition. Thus, our results clearly answer the two questions raised in the Introduction. Apart from general theoretical implications in transport theory, our findings may have experimental relevance as well because researchers usually study how the thermal conductivity changes with the magnetic field for achieving the purpose of understanding and controlling thermal transport.\par

\begin{acknowledgments}
We acknowledge support by the National Natural Science Foundation of China (Grant No. 12105049) and the Natural Science Foundation of Fujian Province (Grant No. 2023J05100).
SL acknowledges the kind hospitality
of Xiamen University where this work was initiated.
\end{acknowledgments}

\end{document}